# Principle of Superposition in Protein-Folding [*]

Satoshi NAKAMURA and Osamu NARIKIYO

*Department of Physics, Kyushu University,*

*Hakozaki, Fukuoka 812-8581*

We propose a principle of superposition to describe protein-folding nuclei and explain an evolutional conservation of a field pattern specifying native-state structure.

In the process of protein folding the dynamics is heterogeneous and folding nuclei play a crucial role. Such folding nuclei are identified, for example, by $\phi$-value analysis in protein engineering experiments. [1] After recent intensive discussions it has been concluded that the sequence of amino acids at the folding nucleus is not evolutionally conserved. [2] In other words, the amino acid at the folding nucleus is variable even in the group of proteins which have almost the same structure and are expected to have the same folding nucleus. Then we try to find an evolutionally conserved quantity, other than the sequence, specifying folding nuclei in this paper.

As a case study we focus our attention to a small protein (TNfn3, PDB code: 1ten), since its folding nuclei are exhaustively studied. [3] The data of the amino-acid sequence and the native state structure is taken from the Protein Data Bank (PDB). Our mini-protein, 1ten, consists of 7 β-sheets and there are 6 folding nuclei, [3] (the type of the amino acid; the residue number in this paper, the residue number in PDB): (ILE; 20, 821), (TYR; 36, 837), (ILE; 48, 849), (LEU; 50, 851), (ILE; 59, 860) and (VAL; 70, 871), on the β-sheets. The residue number is the ordinal number of the amino acid along the sequence.

Since folding nuclei are not easily decided from 1-dimensional (1D) information of the sequence, [2] we consider 3-dimensional (3D) information of the structure. A similar situation has been encountered in the study of the helix-turn-helix motif and it has been shown that 3D keynote, [4] which consists of the list of the number of interactions between pairs of amino acids, can characterize the structure of the motif. The interaction is assigned to each pairs of atoms, excluding hydrogen, if the diameter of the pair is smaller than 6 Å. Such a list reflects 3D information of the structure. We have made the 3D keynote for 1ten but it has not worked well for specifying the folding nuclei. Then in the following we modify the 3D keynote to be suited for describing protein folding.

In the case of protein folding in general it has been recognized for a long time that the formation of hydrophobic core plays an essential role.[5] However, this entropic effect is not taken into account in the analysis of 3D keynote. Thus we introduce a 3D hydrophobicity in order to examine the hydrophobic effect. The 3D hydrophobicity for an amino acid is defined as the sum of the values of hydrophobicity[6] of amino acids within 12 Å from it. This 3D hydrophobicity reflects the structure of the native state. The structure is decided by interactions, including entropic effect, among amino acids and the folding nuclei are related to the formation of hydrophobic core. In Fig. 1(a) the 3D- hydrophobicity profile (HP) is shown as a function of the sequence. The 3D hydrophobicity takes large value at folding nucleus so that it can specify the folding nuclei.

In order to see the evolutional property of the 3D-HP we compare several proteins which have similar structure to 1ten. The similarity of the structure is analyzed by the HSSP server. We plot several profiles of the proteins with the highest Z-scores in HSSP compared with 1ten in Fig. 1(b). From this plot it can be concluded that the 3D-HP is common among these proteins and evolutionally conserved.

In our study we have focused our attention to the folding nuclei on secondary structures, β-sheets, in consistent with the *S*-value analysis.[7] Here we have assumed a hierarchical picture of the folding where the weak residual interactions determine the 3D structure among secondary structures, while the stronger interactions, for example hydrogen bonding, lead to secondary structures before the process of 3D structure formation.

We have shown that the 3D-HP is evolutionally conserved. Then we examine the correlation between the 3D-HP and the $\phi$-value at the folding nuclei. An amino acid with large $\phi$-value plays a role of the folding nucleus.[1] We see a tendency that the larger $\phi$-values are observed at the middle part of the amino-acid sequence: (the residue number; the $\phi$-value,[3] the 3D hydrophobicity): (20; 0.38, 23.77), (36; 0.56, 18.27), (48; 0.67, 14.92), (50; 0.42, 12.66), (59; 0.62, 16.71) and (70; 0.54, 27.37). Thus we consider the buriedness of amino acids in the native state measured by the contact distance. The contact distance is the difference in residue numbers, along the amino-acid sequence, between two amino acids contacting in the native state structure. The amino acids with small contact distance are expected to fold rapidly. The condition for contacts is the same as the above employed 12 Å rule. In Fig. 2 the correlation between the ratio, $\phi$-value / 3D hydrophobicity, and the contact distance is shown. This correlation reflects the fact that the fast processes of the folding correspond to the formation of native contacts at short distance in 1D amino-acid sequence and the slow processes to those at long distance.

In conclusion we have found that the 3D-HP is essential to specify the structure in the native state and evolutionally conserved. This 3D-HP is determined by a superposition of interactions, including entropic effect, among amino acids. The 3D-HP correlates with experimentally measured $\phi$-value after correcting the effect of the contact distance.

Since the 3D-HP is a physico-chemical property superposed over the 3D structure of the native state, the information of the 1D amino-acid sequence is rather unimportant. Thus the structure of proteins is evolutionarily conserved, even if the sequence is not conserved, in accordance with experimental finding. This is a key concept to understand the correspondence between sequence and structure of proteins and illustrated in Fig. 3 where the 3D-HP, a field pattern, is determined by the interaction among amino acids and a specific pattern can be realized by plural different sequences. Our present study is an example of a field theory for proteins.

In our study it is clarified that the field superposed over the native state structure, the 3D-HP, is optimized and evolutionarily conserved. Fig. 4 also supports the cruciality of the superposition. In Fig. 4 it is demonstrated that the zigzag pattern in hydrophobicity along the amino-acid sequence, which leads to β-sheet structure, becomes more evident in the superposition (b) than in the original value (a) of the amino acid. Thus the property of the amino acid itself is rather unimportant but the superposed field realized in the native state structure after interacting with the other amino acids and surrounding water molecules is important.

Our strategy is superposition and distinguished from coarse-graining, e.g. employed in the study of DNA sequence [8] or protein sequence. [9] In the latter study [9] the interpretation of the $\phi$-value on the basis of the sequence was undertaken but failed. [2] Since the experimental importance of the $\phi$-value is obvious, [1] it was embarrassing. Our present study clarifies the significance of the $\phi$-value and resolves the embarrassing situation.

In this paper we have done a case study for 1ten, since it is simple and thoroughly studied mini-protein. In future the same analysis done in this paper should be applied to other proteins and our scenario should be tested for wide class of proteins.

**Figure Captions**

Fig. 1.

The 3D hydrophobicity profile as a function of the residue number of the amino-acid sequence. The 3D hydrophobicity for an amino acid is defined as the sum of the values of hydrophobicity of amino acids within 12 Å from it in the native state structure. The distance between amino acids is determined by that for $C_\beta$'s. In the case of GLY the position of $C_\alpha$ is employed exceptionally. The hydrophobicity for each amino acid is taken from Ref. 6. (a) The 3D hydrophobicity profile for 1ten which consists of 7 β-sheets represented as the horizontal segments (upper), [3] while in PDB assignment 8 segments (lower) exist. (b) The 3D hydrophobicity profiles for 1ten (●) and 3 proteins, 1qr4A (▲), 1fnhA (▼) and 2mfn (■), with the highest Z-scores, $Z>14.5$, in HSSP. The protein 1fnf with $Z=16.6$ is not included in this analysis, since it has many gapped regions in sequence alignment. Six circled data points correspond to 6 folding nuclei described in the text.

Fig. 2.

The correlation between the ratio, $\phi$-value / 3D hydrophobicity, and the sum of the contact distances for the folding nuclei. The contact distance is the difference in residue numbers between two amino acids contacting in the native state structure. The condition for contacts is the same as that employed in Fig. 1.

Fig. 3.

A field pattern can be realized by plural different sequences.

Fig. 4

(a) The hydrophobicity for each amino acid as a function of the residue number of the amino-acid sequence. (b) The 3D hydrophobicity profile with the directional weight. The 3D hydrophobicity for an amino acid in this case is defined in a similar manner as in Fig. 1 but in the calculation of the sum the directional weight is multiplied where the weight is +1 for the amino acids whose side chain is in the inside of 1ten and −1 for outside. This rule takes into account the fact that the positive hydrophobicity is preferred inside but negative one is preferred outside.

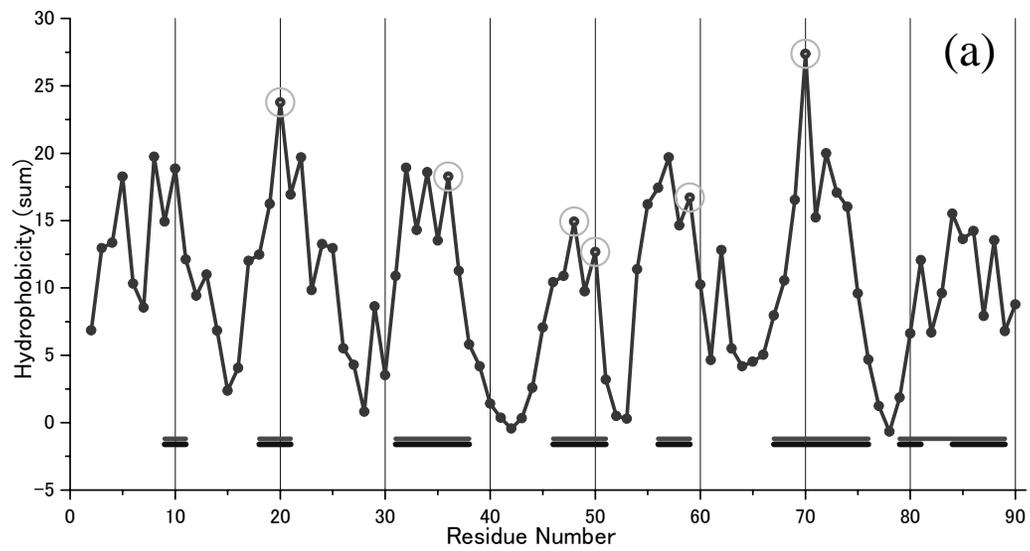

Fig. 1 (a)

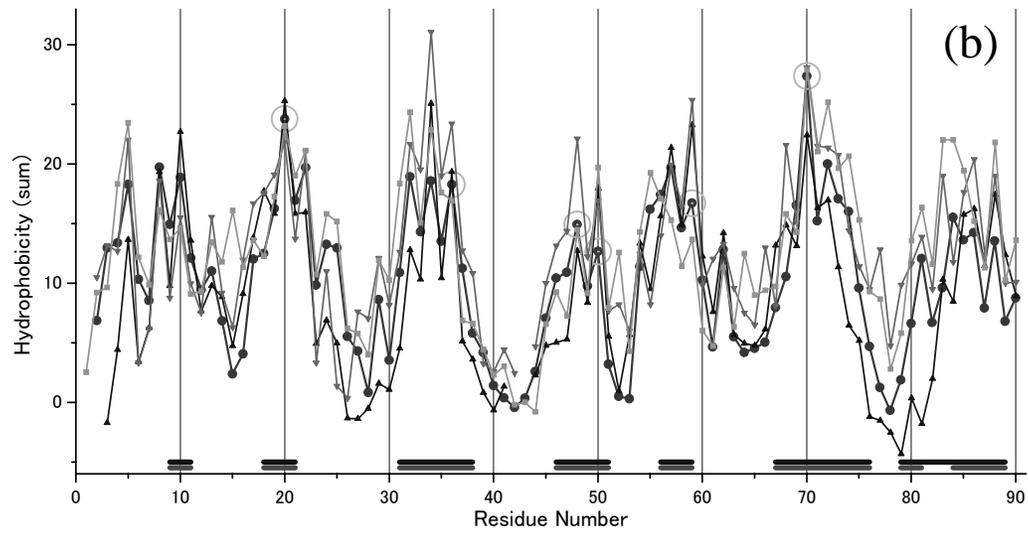

Fig. 1 (b)

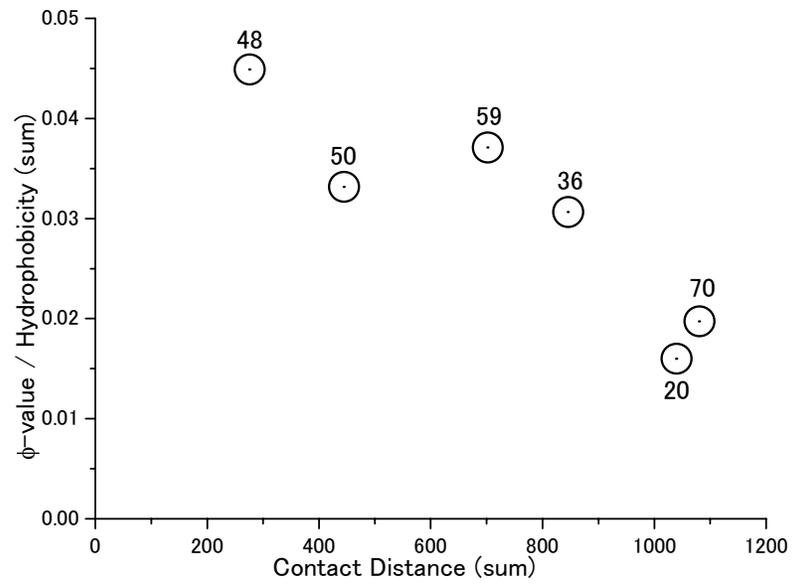

Fig. 2

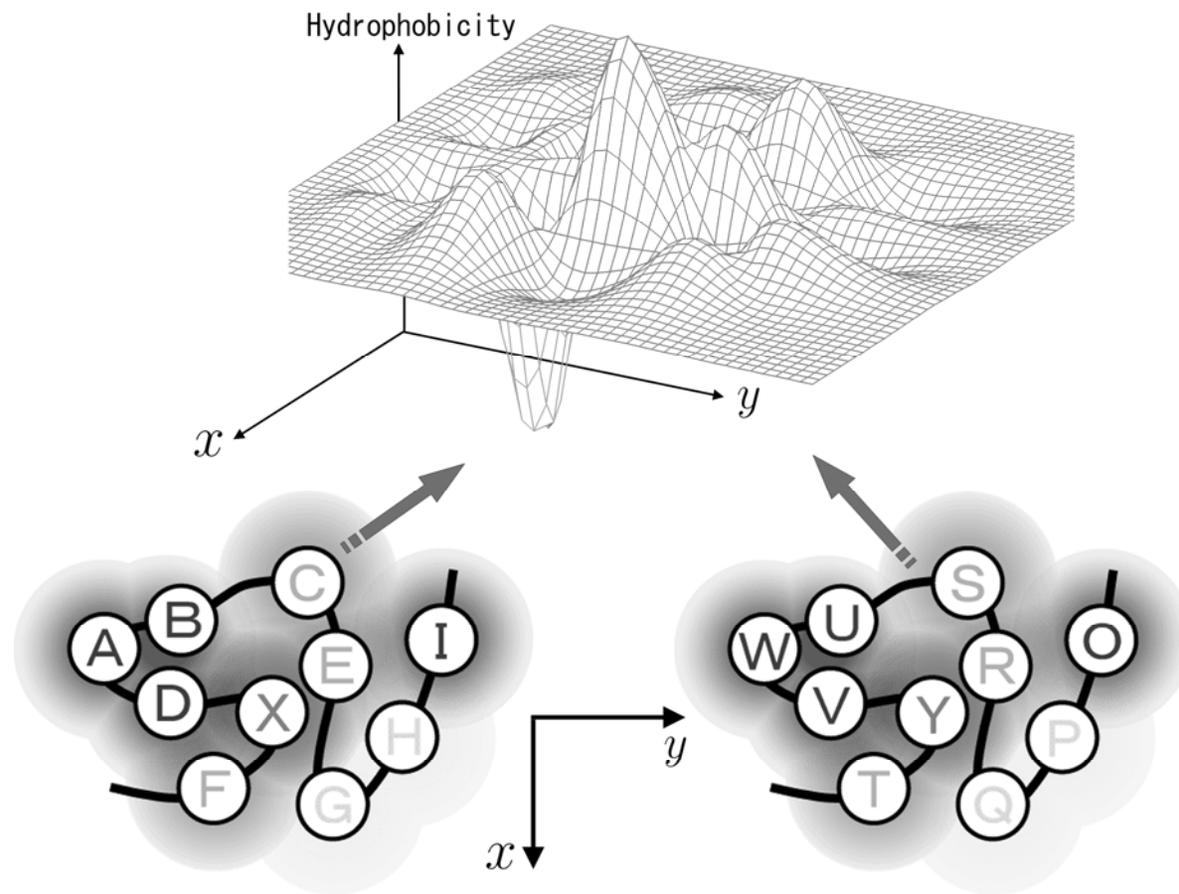

Fig. 3

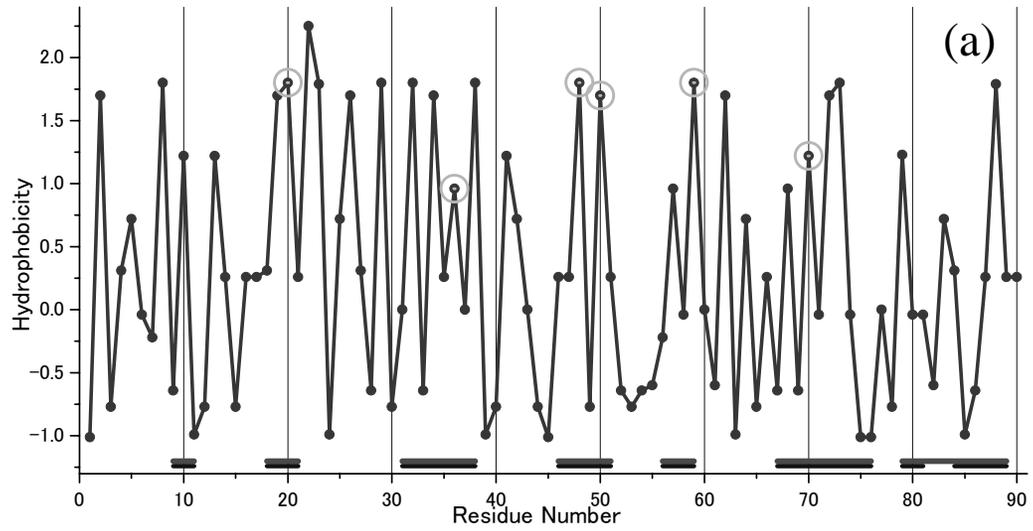

Fig. 4 (a)

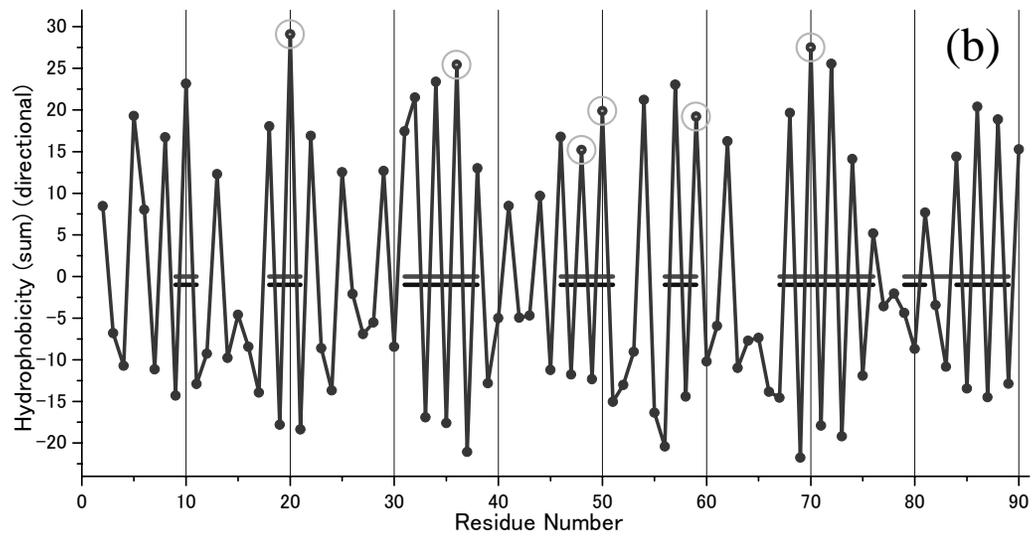

Fig. 4 (b)